\documentclass[acmsmall,screen]{acmart}

\usepackage{soul}
\usepackage{threeparttable}
\usepackage[inline]{enumitem}    
\usepackage[greek,english]{babel}
\usepackage{ifthen}
\usepackage{xspace}
\usepackage{fancybox}
\usepackage{marginnote}
\usepackage{multirow}
\usepackage{mathtools}
\usepackage{hyperref}
\hypersetup{
    colorlinks=true,
    linkcolor=blue,
    filecolor=magenta,      
    urlcolor=cyan,
    pdftitle={Overleaf Example},
    pdfpagemode=FullScreen,
    }

\usepackage[noend]{algpseudocode}
\usepackage{graphicx}
\usepackage{cleveref}
\usepackage{balance}
\usepackage{makecell}
\usepackage{graphicx}
\usepackage{algorithm}
\usepackage{minted}
\usepackage{multirow}
\usepackage{float}
\usepackage{booktabs}
\usepackage[utf8]{inputenc}
\usepackage{dirtytalk}
\usepackage{epigraph}
\usepackage{latexsym}
\usepackage{tikz}
\usepackage{colortbl}
\usepackage{tabulary}
\usepackage{etoolbox}
\usepackage{asymptote}
\usepackage{amsmath}
\usepackage{graphicx}
\usetikzlibrary{calc,fit,positioning,shapes.misc,shapes.geometric,shapes,arrows,chains,mindmap,trees,backgrounds, patterns}
\usepackage{listings}
\usepackage{booktabs,caption}
\usepackage{multicol}
\usepackage{lipsum}% just to generate filler text

\usepackage{subcaption}
\usepackage{wrapfig}
\usepackage{array}
\usepackage{collcell}
\usepackage{pgfplots, pgfplotstable}
\pgfplotsset{compat=newest}
\usepackage{fontawesome5}
\usepackage{xcolor}

\usepackage[normalem]{ulem}

\AtBeginEnvironment{grammar}{\small}

\newboolean{reviews} 
\setboolean{reviews}{false}
\ifthenelse{\boolean{reviews}}{

}
{

}

\newboolean{showcomments} 
\setboolean{showcomments}{true}
\ifthenelse{\boolean{showcomments}}
{\newcommand{\nb}[2]{
		\fbox{\bfseries\sffamily\scriptsize#1}
		{\sf\small$\blacktriangleright$\textit{\textcolor{red}{#2}}$\blacktriangleleft$}
	}
}
{\newcommand{\nb}[2]{}}

\ifthenelse{\boolean{showcomments}}
{
}
{}

\crefname{lstlisting}{listing}{listings}
\usepackage{fp}

\newcommand{\code}[1]{\texttt{\fontsize{8.5}{10}\selectfont #1}}

\newcommand{\py}{\texttt{Python}\xspace}

\newcommand{\pycrypto}{\texttt{pycryptodome}\xspace}
\newcommand{\crypto}{\texttt{cryptography}\xspace}

\algnewcommand\algorithmicforeach{\textbf{for each}}
\algdef{S}[FOR]{ForEach}[1]{\algorithmicforeach\ #1\ \algorithmicdo}

\newcommand{\lstbg}[3][0pt]{{\fboxsep#1
  \colorbox{#2}{\strut \ensuremath{#3}}}}

\lstdefinelanguage{diff}{
  morecomment=[f][\color{red}]-,
  morecomment=[f][\color{green!50!black}]+,
}

\lstdefinelanguage{smdiff}{
  basicstyle=\ttfamily\scriptsize,
  morecomment=[f][\lstbg{red!20}]-,
  morecomment=[f][\lstbg{green!20}]+
}

\lstdefinelanguage{diff_1}{
  basicstyle=\ttfamily\tiny,
  morecomment=[f][\lstbg{red!20}]-,
  morecomment=[f][\lstbg{green!20}]+,
  morecomment=[l][\color{orange}]{\ \ private\ IUIModes\ um}
}

\usepackage{etoolbox}
\usepackage{fvextra}
\patchcmd{\theFancyVerbLine}{\arabic{FancyVerbLine}}{\sffamily\footnotesize\arabic{FancyVerbLine}}{}{}

\definecolor{red_diff}{RGB}{255,204,203}
\definecolor{dkgreen}{rgb}{0,0.6,0}
\definecolor{gray}{rgb}{0.5,0.5,0.5}
\definecolor{mauve}{rgb}{0.58,0,0.82}
\definecolor{lightblue}{rgb}{0.8,0.93,1}

\iftrue
    \newcommand\daniel[1]{\textcolor{blue}{Daniel: #1}}
    \newcommand{\clg}[1]{\textcolor{red}{CLG: #1}}
    \newcommand{\ruben}[1]{\textcolor{magenta}{RM: #1}}
\else
    \newcommand\daniel[1]{}
    \newcommand{\clg}[1]{}
    \newcommand{\ruben}[1]{}
\fi

\newcommand{\piranha}{\texttt{PolyglotPiranha}\xspace}
\newcommand{\tool}{\textsc{Spell}\xspace}
\newcommand{\melt}{\textsc{Melt}\xspace}
\newcommand{\ideas}{use cases\xspace}
\newcommand{\idea}{use case\xspace}

\usepackage{xspace}

\usepackage{etoolbox} % for patching commands
\let\oldttfamily\ttfamily % hacks for nice fonts.
\renewcommand{\ttfamily}{\oldttfamily\footnotesize}
\renewcommand{\code}[1]{\lstinline|#1|}

\definecolor{darkblue}{HTML}{1F77B4}
\definecolor{darkorange}{HTML}{FF7F0E}
\definecolor{darkgreen}{HTML}{006400}
\definecolor{darkred}{HTML}{D62728}
\definecolor{darkpurple}{HTML}{9467BD}
\definecolor{darkbrown}{HTML}{8C564B}
\definecolor{darkpink}{HTML}{E377C2}
\definecolor{darkgray}{HTML}{7F7F7F}
\definecolor{darkcyan}{HTML}{17BECF}

\lstdefinelanguage{langs}{
  moredelim=[is][\color{darkgreen}]{*}{*},
  moredelim=*[s][\color{black}]{@}{\ },
  moredelim=*[s][\color{black}]{(}{)},
  morekeywords={},
}

\usepackage[svgnames]{xcolor}
  \definecolor{diffstart}{named}{Grey}
  \definecolor{diffincl}{named}{Green}
  \definecolor{diffrem}{named}{OrangeRed}

\usepackage{listings}
  \lstdefinelanguage{diff}{
    basicstyle=\ttfamily\small,
    morecomment=[f][\color{diffstart}]{@@},
    morecomment=[f][\color{diffincl}]{+\ },
    morecomment=[f][\color{diffrem}]{-\ },
  }

\usepackage{amsthm}
\usepackage{thmtools}
\usepackage[most]{tcolorbox}
\usepackage[nounderscore]{syntax}

\usepackage{DejaVuSansMono}
\usepackage[T1]{fontenc}
\definecolor{highlightblue}{RGB}{33, 111, 199}

\newtcolorbox[auto counter, number within=section]{highlight}[1][]{
  colback=highlightblue!5!white,
  colframe=highlightblue!80!black,
  coltitle=highlightblue!90!black,
  colbacktitle=highlightblue!15!white,
  fonttitle=\bfseries\footnotesize,
  title=Finding~\thetcbcounter,
  boxrule=0.8pt,
  arc=4pt,
  outer arc=4pt,
  boxsep=4pt,
  left=3pt,
  right=3pt,
  top=3pt,
  bottom=3pt,
  toptitle=2pt,
  bottomtitle=2pt,
  titlerule=0pt,
  attach boxed title to top left={yshift=-\tcboxedtitleheight/2, xshift=2mm},
  boxed title style={arc=3pt, outer arc=3pt},
  #1
}

\begin{document}

\title{\textsc{Spell}: Synthesis of Programmatic Edits using LLMs}

\author{Daniel Ramos}
\email{danielrr@cmu.edu}
\orcid{0000-0002-2147-2176}
\affiliation{%
  \institution{Carnegie Mellon University, and INESC-ID / IST - Universidade de Lisboa}
  \country{Portugal}
}

\author{Catarina Gamboa}
\email{cgamboa@andrew.cmu.edu}
\orcid{0000-0002-6995-7340}
\affiliation{%
  \institution{Carnegie Mellon University, and LASIGE / FCUL - Universidade de Lisboa}
  \country{Portugal}
}

\author{In\^es Lynce}
\email{ines.lynce@inesc-id.pt}
\orcid{0000-0003-4868-415X}
\affiliation{
  \institution{INESC-ID / IST - Universidade de Lisboa}
  \country{Portugal}
}

\author{Vasco Manquinho}
\email{vasco.maquinho@inesc-id.pt}
\orcid{0000-0002-4205-2189}
\affiliation{
  \institution{INESC-ID / IST - Universidade de Lisboa}
  \country{Portugal}
}

\author{Ruben Martins}
\email{rubenm@cs.cmu.edu}
\orcid{0000-0003-1525-1382}
\affiliation{%
  \institution{Carnegie Mellon University}
  \country{USA}
}

\author{Claire Le~Goues}
\email{clegoues@cs.cmu.edu}
\orcid{0000-0002-3931-060X}
\affiliation{%
  \institution{Carnegie Mellon University}
  \country{USA}
}

\renewcommand{\shortauthors}{Ramos et al.}

\begin{abstract}

Library migration is a common but error-prone task in software development. Developers may need to replace one library with another due to reasons like changing requirements or licensing changes. Migration typically entails updating and rewriting source code manually. While automated migration tools exist, most rely on mining examples from real-world projects that have already undergone similar migrations. However, these data are scarce, and collecting them for arbitrary pairs of libraries is difficult. Moreover, these migration tools often miss out on leveraging modern code transformation infrastructure.

In this paper, we present a new approach to automated API migration that sidesteps the limitations described above. Instead of relying on existing migration data or using LLMs directly for transformation, we use LLMs to extract migration examples. Next, we use an Agent to generalize those examples to reusable transformation scripts in \piranha, a modern code transformation tool. Our method distills latent migration knowledge from LLMs into structured, testable, and repeatable migration logic, without requiring preexisting corpora or manual engineering effort. Experimental results across \py libraries show that our system can generate diverse migration examples and synthesize transformation scripts that generalize to real-world codebases.

\end{abstract}

\setcopyright{none} % to remove the copyright notice
\settopmatter{printacmref=false} % to remove the ACM Reference Format
\renewcommand\footnotetextcopyrightpermission[1]{}

\maketitle

\section{Introduction}

\begin{figure*}[t]
  \centering
  \begin{subfigure}[t]{0.48\textwidth}
    \centering
    \caption{Before (using \texttt{cryptography.fernet})}
    \label{fig:encrypt-func-before}
    \begin{lstlisting}[
          language=Python,
          basicstyle=\scriptsize\oldttfamily,
          numbers=left,
          numberstyle=\scriptsize,
          numbersep=5pt,
          escapeinside={*@}{@*}
        ]
import hashlib
from cryptography.fernet import Fernet
(...)
def encrypt_document(document: str, key: bytes) -> bytes: *@\label{line:encrypt-before}@*
    cipher = Fernet(key)
    encrypted = cipher.encrypt(document.encode())
    return encrypted
(...)
    \end{lstlisting}
  \end{subfigure}
  \hfill
  \begin{subfigure}[t]{0.48\textwidth}
    \centering
    \caption{After (using \texttt{pycryptodome})}
    \label{fig:encrypt-func-after}
    \begin{lstlisting}[
      language=Python,
      basicstyle=\scriptsize\oldttfamily,
      numbers=left,
      numberstyle=\scriptsize,
      numbersep=5pt,
      escapeinside={*@}{@*}
    ]
from Crypto.Cipher import AES
from Crypto.Util.Padding import pad
(...)
def encrypt_document(document: str, key: bytes) -> bytes: *@\label{line:encrypt-after}@*
    cipher = AES.new(key, AES.MODE_CBC)
    padded_data = pad(document.encode(), AES.block_size)
    encrypted = iv + cipher.encrypt(padded_data)
    return encrypted
    \end{lstlisting}
  \end{subfigure}
  
  \vspace{0.5em}
  
  \begin{subfigure}[t]{\linewidth}
    \centering
    \caption{Synthesized \piranha program expressing the transformation}
    \label{fig:script-gen}
    \includegraphics[width=\linewidth]{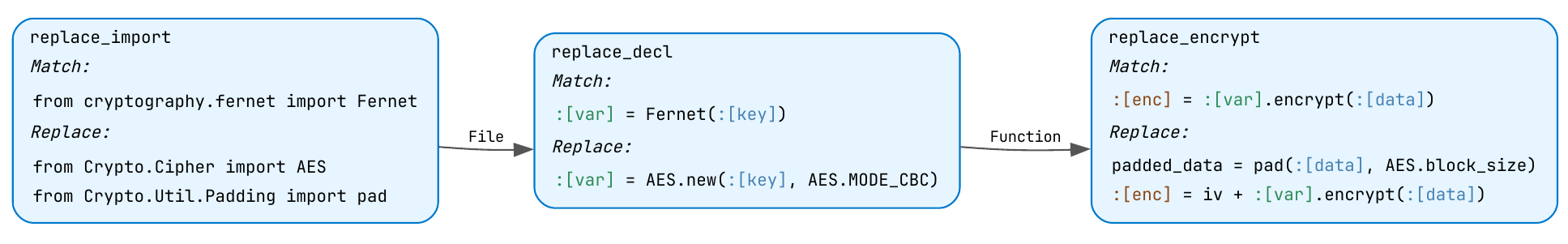}
  \end{subfigure}
  \caption{\small Migration from \texttt{cryptography.fernet} to \texttt{pycryptodome}: function-level encryption logic before and after migration, with the corresponding \piranha transformation script extracted from this example.}
  \label{fig:encrypt-migration-complete}
\end{figure*}

Libraries are the cornerstone of modern software development. They provide modular functionality through their application programming interfaces (APIs)~\cite{de2004good}, which expose functionality while abstracting implementation details. 
Library \emph{migration} refers to the process of replacing one software library used in a client codebase with another. Such migrations are necessary to accommodate evolving software requirements, e.g., a need for new features, license changes, or more modern or better-supported APIs~\cite{DBLP:conf/wcre/XavierBHV17, DBLP:conf/sigsoft/BogartKHT16, DBLP:journals/smr/DigJ06, chapin2001types, perkins2005automatically}.
Replacing an original library’s API calls with those of the target library is a largely manual~\cite{DBLP:conf/sigsoft/KimZN12} and error-prone~\cite{DBLP:conf/icse/Kim0DB18} task.

The widespread and costly nature of library migrations motivates research into automation. Most existing methods mine software repositories that have already undergone migration~\cite{meditor, soar, apifix, apimigrator, a3, appevolve, tcinfer, inferrule}, generalizing and systematizing examples into reusable formats.
However, these techniques are typically limited by their dependency on the availability and quality of historical migration data, which is scarce in practice~\cite{developers_dont_update}. 

\looseness-1
Meanwhile, large language models (LLMs) have demonstrated potential and gained traction in assisting with code transformation, including in industrial settings (e.g., Amazon Q Transform~\cite{AmazonQTransform}). However, directly using LLMs for migration is often unreliable, introducing unpredictable or formatting/stylistic changes or subtle semantic bugs~\cite{jesse2023large}. They can also be inconsistent across similar migration patterns, and provide no reusable or inspectable artifacts for developers to understand what has been done.  Finally, LLM-based migrations must inefficiently generate most changes from scratch.  

In this paper, we introduce \tool (\textbf{S}ynthesis of \textbf{P}rogrammatic \textbf{E}dits using \textbf{LL}Ms), which, rather than using LLMs as black-box code translators, uses them as \emph{knowledge sources} to systematically generate validated migration examples that are then distilled into precise, reusable transformation scripts. Through pretraining on large-scale code corpora, LLMs learn joint representations of APIs and the semantic relationships between them, including cross-library mappings~\cite{transcoder}.  That is, LLMs implicitly encode rich knowledge about API relationships and usage patterns.
This knowledge may be imperfect, or too noisy for direct application. However, our \emph{key insight} is that it can be systematically extracted and validated to support migration script synthesis. 
This combines broad API knowledge encoded in LLMs with the reliability and efficiency of rule-based transformations, without requiring historical migration data. 

A key benefit of \tool is that it generates migration scripts in a domain-specific language (DSL) designed for code transformation (specifically \piranha~\cite{piranha_pldi}). While most prior work produces or relies on custom transformation toolsets~\cite{apifix, apimigrator, inferrule, appevolve}, \tool aims to leverage modern DSLs and frameworks like OpenRewrite~\cite{OpenRewrite}, GritQL~\cite{GritQL}, and \piranha~\cite{piranha_pldi} that support automated, consistent transformations across large codebases.
\tool's output thus serves as maintainable assets that can be inspected, version-controlled, debugged, and refined by developers. These DSLs are increasingly integrated into CI/CD pipelines at major technology companies like Uber~\cite{piranha_pldi} and Amazon~\cite{AmazonQTransform}, enabling our approach to naturally extend modern workflows.
The overall use case can benefit both library maintainers supporting client codebases when pushing deprecating changes, and client teams migrating code without historical data available to mine. 

\tool takes a novel, hybrid approach to synthesizing migration scripts in the \piranha language from its generated examples.  
\tool first uses a classic approach for anti-unification-based synthesis.  This step deterministically infers syntactic patterns from concrete code differences to produce an initial set of local transformation rules that express parts of a migration. 
However, \piranha, like other transformation frameworks and DSLs~\cite{coccinelle,OpenRewrite,GritQL}, 
expresses migrations as combinations of locally-scoped rules, to capture semantic relationships, dependencies, chaining, or scoping. 
\tool therefore combines classic synthesis with an agent workflow to refine, compose, and orchestrate the initially-inferred local rules into overall \piranha programs that take full advantage of the expressive power of modern transformation DSLs.  

%, using an agent workflow. This allows us to produce reusable, %testable scripts without manual intervention or large mined datasets. %In particular, our approach consists of two main steps. First, we extract migration knowledge from LLMs in the form of migration examples. Second, we generalize these into migration scripts using a code transformation toolset.

%Overall, \tool is a novel method for automatically generating reusable, testable, validated migration scripts that leverage an existing, powerful transformation toolset. It does not require historical migration data, and leverages and tests latent LLM knowledge without relying entirely on them for vanilla migration (avoiding associated downsides).  
Our results show that \tool can synthesize correct, reusable migration scripts across diverse Python library migrations. For nine migration tasks, it generated an average of 87 validated examples per task and inferred correct scripts for 61.6\% in a single trial. \tool outperformed \melt, a state-of-the-art prior approach,  on every task. Moreover, the scripts generalized to real-world projects taken from open-source repositories. We kept our LLM budget deliberately low ($ < \$100$) for all experiments, demonstrating that the results are attainable cost-effectively, using small models. These findings show that structured migration knowledge can be distilled from LLMs and converted into reusable migration scripts.

In summary, we make the following contributions:
\begin{enumerate}
\item A novel method for automatically generating reusable, validated migration scripts in an existing transformation DSL. 
\item A data generation approach for systematically extracting and validating structured migration knowledge from LLMs without requiring historical migration logs.
\item A novel agentic workflow that leverages classic synthesis to transform examples into reusable migration scripts.
\item A new dataset of 870 validated migration examples with corresponding test cases across ten \py migration tasks.
\item An implementation, \tool, which achieves a 61.6\% correctness rate in one-shot script synthesis and outperforms the previous state-of-the-art on every evaluated task. We demonstrate the application of these scripts to real, open-source projects. 
\item An open-source artifact including the \tool implementation, dataset, evaluation harness, and all synthesized scripts and results, publicly available~\cite{artifact}.
\end{enumerate}

\section{Illustrative Example}
\label{sec:motivation}

\begin{figure*}[t]
    \centering
    \includegraphics[width=1\linewidth]{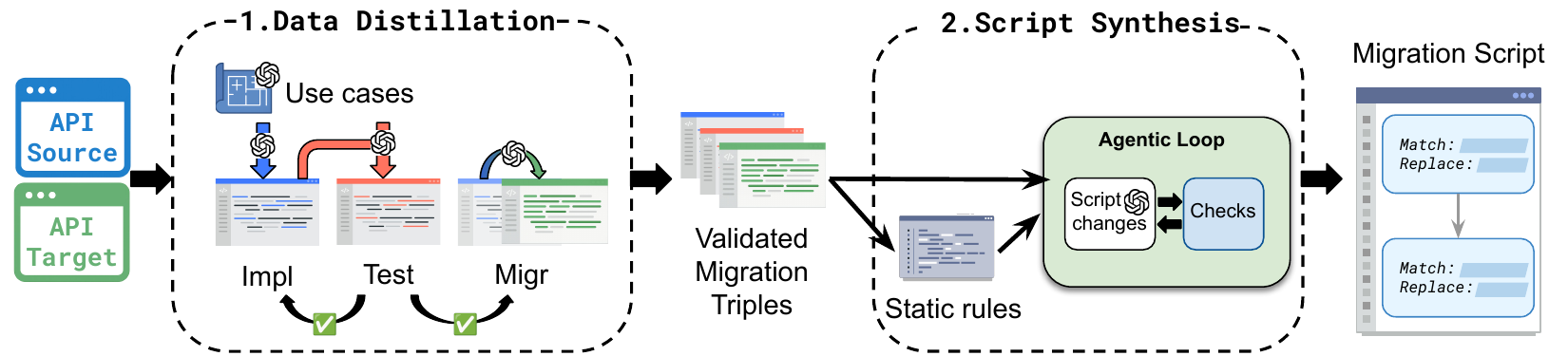}
    \caption{\small Overview of \tool's pipeline. Given a pair of source and target APIs, Phase 1 distills data from LLMs in the form of use cases. These use cases guide the generation of examples that contain an implementation in the source API, a migration to the target API, and tests that validate both implementations. Phase 2 synthesizes the migration script using an initial set of statically inferred rules and an agentic loop to update them. The final output is a project-agnostic migration script that can be automatically applied to migrate from a source to a target API.}
    \label{fig:metadiagram}
\end{figure*}

Our pipeline consists of two phases: (1) data distillation, which generates code migration examples with tests, and (2) synthesis, where we abstract those pairs into reusable migration scripts. \Cref{fig:metadiagram} depicts the overall pipeline.

\vspace{1ex}
\noindent\textbf{Data Distillation.}
The goal of data generation is to distill code migration knowledge from the LLM in the form of examples.
To illustrate, consider the task of migrating between two Python cryptography libraries: \crypto\footnote{https://github.com/pyca/cryptography} and \pycrypto.\footnote{https://github.com/Legrandin/pycryptodome} Given a pair of libraries \textit{(source, target)}, our approach uses an LLM to generate two implementations of the same functionality, one using each library. These paired implementations form the basis for migration examples that are later generalized.

The first step asks an LLM to generate \emph{\ideas}, i.e., tasks or features that could plausibly be implemented using either library. These \ideas help focus the model on a shared functional goal that can be realized with both APIs. When asked to propose such a \idea for \crypto and \pycrypto, \texttt{gpt4o-mini} responds:

\newenvironment{smallquote}
  {\list{}{\leftmargin=1.5em\rightmargin=1em}\item[]}
  {\endlist}
  
\begin{smallquote} 
\emph{
A \textbf{File Encrypting Utility} can be implemented using \pycrypto or \crypto to allow users to encrypt and decrypt textual documents. High-level behaviour and API: (...)} \end{smallquote}

We then prompt the model to generate an \emph{implementation} for this \idea using the source library. \Cref{fig:encrypt-func-before} shows one such implementation generated by the LLM using the \crypto~library. 
We also prompt the LLM to generate \emph{tests} for each implementation that verify functionality, and filter out clearly-spurious examples. %\ruben{It is unclear how these are filtered out.}

Finally, we prompt the LLM to \emph{migrate} each implementation in \crypto to the alternative library \pycrypto. \Cref{fig:encrypt-func-after} shows a migration generated for the code in \Cref{fig:encrypt-func-before}. 
The migration prompt instructs the LLM to modify only the library-specific API calls, leaving the rest of the code unchanged. This enables the reuse of the tests generated for the original implementation on the migrated version.  The tests provide a signal of migration success when the migrated code's tested behavior is the same as the original, and thus serve as a filter on the examples. 
%\ruben{We could talk about coverage here to give some information on the quality of tests.}

The process produces a set of code migration examples consisting of pairs of implementations of a \idea, with shared test files. \Cref{sec:data} describes data generation in detail. 

\vspace{1ex}
\noindent\textbf{Synthesizing Migration Scripts.}
%\label{sec:piranha-intro}
%
The second phase abstracts synthetic migration examples into reusable migration programs that capture structural transformation patterns and can be applied across similar tasks or large codebases. 
We generate programs in the \piranha~\cite{piranha_pldi} language, a DSL for source-to-source transformations. A \piranha program is composed of \texttt{rules}, each with a \texttt{match} clause that identifies a code pattern and a \texttt{replace} clause that rewrites it. Template variables (e.g., \code{:[var]}) allow rules to generalize over specific names and expressions, acting like regular expressions, but for syntax-aware code patterns.
Rules can be connected through labeled edges that specify how follow-up transformations should be triggered. This enables cascading and context-sensitive rewrites.

\Cref{fig:script-gen} shows a \piranha program synthesized from the example.  The first rule, \texttt{replace import}, replaces the original import with one from the target library. This triggers the second rule within the same file, \texttt{replace decl}, which rewrites the object construction. The third rule, \texttt{replace encrypt}, updates how the encryption API is used. Template variables \texttt{:[var]} and \texttt{:[data]} allow the rules to generalize to different programs that share the same structure, but use different variable names or concrete expressions. 

%In \Cref{sec:synth}, we describe how our agent workflow automatically synthesizes these rules from the migration examples. 
The output of this step is a set of interpretable migration programs that capture the essence of the transformation and can be adapted or extended to similar scenarios. \Cref{sec:synth} details script synthesis from synthetic examples.

\section{Migration Data Distillation}
\label{sec:data}

\begin{figure*}[t]
    \centering
    \includegraphics[width=1\linewidth]{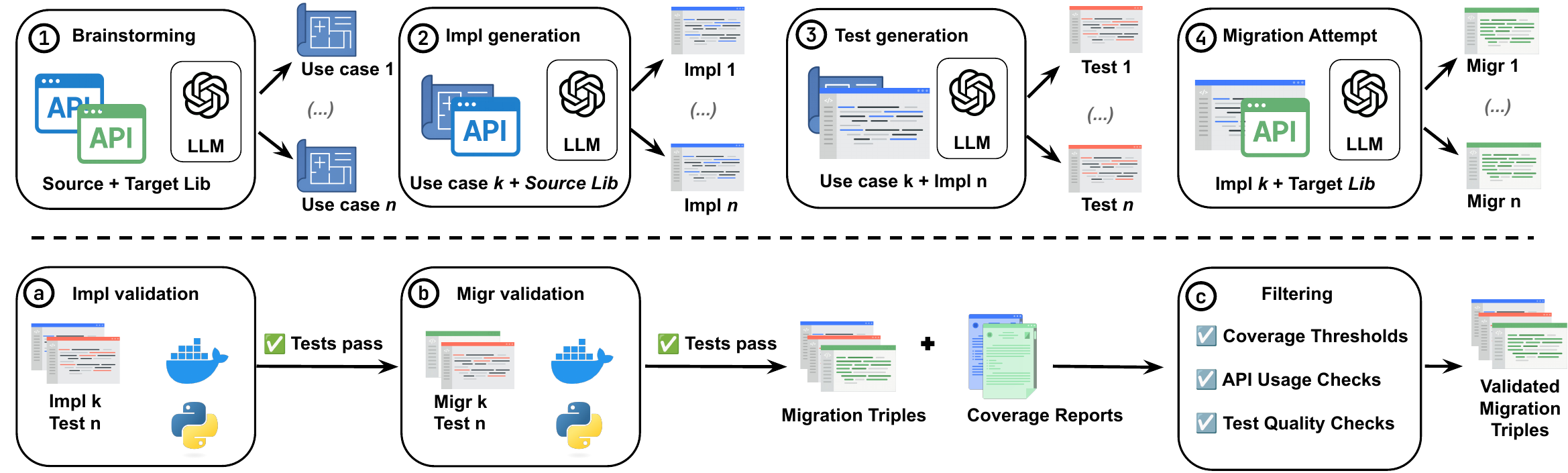}
    \caption{\small Pipeline for generating migration examples using LLMs. Given a source–target library pair, we prompt the LLM to propose abstract \emph{\ideas} leveraging latent knowledge of both libraries (Step 1). For each \idea, we generate multiple source library implementations (Step 2), and test suites (Step 3), and migrate them to the target library (Step 4). Validation entails implementation validation (a), migration validation (b), and quality filtering (c). The output is a collection of validated migration triples (implementation, test, migration).}
    \label{fig:data-gen}
\end{figure*} 

\Cref{fig:data-gen} overviews the data distillation process, which extracts and validates LLM's pre-trained knowledge about migrating between two libraries. The process generates sets of equivalent programs using source and target libraries, with test cases that provide evidence of their functional equivalence.  This data is the input for migration script synthesis (\Cref{sec:synth}). 

More precisely: given source library $\mathcal{S}$ and target library $\mathcal{T}$, we sample migration knowledge from an LLM as raw migration triples $\widetilde{\mathcal{M}} = \{(\tilde{s}_i, \tilde{t}_i, \tilde{m}_i)\}_{i=1}^n$. These triples are automatically generated and may contain redundancy, inconsistencies, or errors. 
We validate and filter this set to produce a high-quality subset $\mathcal{M} = {(s_i, t_i, m_i)}_{i=1}^m$, where $m \leq n$, and:
\begin{itemize}
\item $s_i$ is an implementation using $\mathcal{S}$
\item $m_i$ is a functionally equivalent implementation using  $\mathcal{T}$  
\item $t_i$ is a test suite that executes and passes on both $s_i$ and $m_i$
%\item $\mathcal{M}$ is the subset of validated migration triples extracted from $\widetilde{\mathcal{M}}$
\end{itemize}
%\ruben{It is unclear when we are using the tidle notation; explicitly say what tidle means.}
\looseness-1
We use $\sim$ to denote sampled triples that may or may not be valid; notation without a $\sim$ refers to data/triples that have been validated.
We now explain knowledge extraction (\Cref{sec:sampling}) and validation (\Cref{sec:validation}). Prompts are omitted for space, but are available in our  artifact~\cite{artifact}. 

\subsection{Model Sampling}
\label{sec:sampling}

The goal in model sampling is to generate a diverse initial set of example implementations of shared functionality, using $\mathcal{S}$ and $\mathcal{T}$.
A nontrivial library can usually be used in a variety of ways.  Thus, our process entails a multi-stage pipeline that first generates multiple \emph{\ideas} of functionality that can be implemented using $\mathcal{S}$ (Section~\ref{sec:ideagen}), and separately generating \emph{implementations} of those \ideas using $\mathcal{S}$ (Section~\ref{sec:implementation}).  
For each implementation of each \idea, the pipeline generates tests  (Section~\ref{sec:testgen}), and finally migrates the implementation to use $\mathcal{T}$  (Section~\ref{sec:migration}).  

%This produces a set $\mathcal{M}$ of migration triples, potential example migrations between $\mathcal{S}$ and $\mathcal{T}$, with tests. These examples are validated and filtered as described in \Cref{sec:validation}. 
% CLG I think the above has been beaten to death at this point, so I'm risking leaving out the roadmapping.

\subsubsection{Use Case Generation}
\label{sec:ideagen}

Given libraries $(\mathcal{S}, \mathcal{T})$, \tool first prompts an LLM to generate multiple \emph{\ideas} $\mathcal{C}$ (\Cref{fig:data-gen}, Step 1). Each \idea $C_j \in \mathcal{C}$ represents a use case, program, or otherwise useful functionality that could be implemented using either library.

Generating \ideas first decouples the creative task of brainstorming migration scenarios from the technical task of writing correct code.  This serves to encourage a model to explore a wide space of API usages, beyond common boilerplate. It also provides a clear, shared semantic goal, encouraging consistency between source and target implementations.  Finally, it allows the pipeline to retry implementation generation for promising \ideas as necessary, without having to re-explore the entire conceptual space. Decoupling also allows the model to focus on brainstorming \ideas that are not complex or otherwise unsuitable, before investing computational resources in their implementation, next.

Use case generation begins by using the model to create a small set of seed \ideas that are then used for a self-instruct loop to generate additional migration scenarios.
Self-instruction, in which a model produces its own task demonstrations using seed data, has proven effective in bootstrapping a model's existing knowledge in a variety of contexts~\cite{selfinstruct}.
We first generate $p$ seed \ideas by prompting the model directly with a basic instruction. Each \idea describes a use case involving a source library, along with a high-level summary of its functionality. These seed \ideas are collected into a list and used as few-shot examples in subsequent prompts to generate more.

The \idea generation loop proceeds by randomly sampling $k$ prior \ideas (seed or generated) to include in a new prompt, and asking the model to propose one new \idea. This repeats until it produces a target number of \ideas, where each \idea demonstrates a way the source and target libraries can be used.
%This step outputs a set of $r$ \emph{\ideas} that demonstrate ways the source and target libraries can be used.

\subsubsection{Implementation Generation}
\label{sec:implementation}

For each \idea $C_j \in \mathcal{C}$, we prompt the model to generate a set of implementations $\widetilde{S}_j = \{\tilde{s}_{j,1}, \tilde{s}_{j,2}, ..., \tilde{s}_{j,n}\}$ using source library $\mathcal{S}$ (\Cref{fig:data-gen}, Step 2). 
Ideally, these implementations are modular, with a well-defined API. This simplifies migration to $\mathcal{T}$ while preserving the high-level interface, necessary to use the same tests on both.  

The prompt thus instructs the model to implement the \idea in a self-contained \py file, subject to the following constraints: 
\begin{enumerate*} 
\item implement the API for \idea $C_j$ as a set of functions,
\item focus on API implementation; a small \texttt{main} block is optional, and
\item avoid external dependencies beyond the specified library $\mathcal{S}$. 
\end{enumerate*}

We construct a multi-turn chat prompt that includes the original \idea and implementation instructions. To obtain diverse outputs and increase the chances that the model produces usable implementations, we sample multiple completions using stochastic decoding. 

The output of this step is a set of up to $n$ potential \emph{implementations} of each $C_j \in \mathcal{C}$. 

%The implementations may not be functional or correct; they are validated and filtered as described in Section~\ref{sec:validation}.

\subsubsection{Test Generation}
\label{sec:testgen}

For each \idea-implementation pair $(C_j, \tilde{s}_{j,k})$, we generate $p$ test suites $\widetilde{T}_{j,k} = \{\tilde{t}_{j,k,1}, \tilde{t}_{j,k,2}, ..., \tilde{t}_{j,k,p}\}$ (\Cref{fig:data-gen}, Step 3). 
%A test suite $T_{j,k}$ is intended to partially verify, and build confidence in, the correctness of implementation  $s_{j,k}$ (and, later, the migrated implementation using target library $\mathcal{T})$.
Each test suite $\tilde{t}_{j,k,l}$ is intended to partially verify, and build confidence in the correctness of implementation $\tilde{s}_{j,k}$ (and, later, the migrated implementation using target library $\mathcal{T})$.

The ideal tests target an implementation's public API and core functionality, rather than internal implementation details that may not transfer.
For example, tests for a program that encrypts files should not inspect the encrypted output directly, but should instead verify that decrypting a file restores the original content (among other functionality).
The prompt instructs the model to generate unit tests (or end-to-end tests, if unit tests are infeasible) that validate the overall system behavior.

The prompt includes $C_j$,  $\tilde{s}_{j,k}$, and the test generation instructions.
We sample multiple test files per implementation to mitigate model hallucination as well as challenges in test generation generally, such as incorrect assertions.  
Sampling increases the chances of obtaining at least one valid test file for an otherwise correct implementation. 
Since our implementation is \py-specific, the prompt instructs the model to use either the \texttt{pytest} testing framework or property-based testing with \texttt{hypothesis}, when appropriate.
This assumption could be easily ported to other languages. 

The output of this step is multiple test files per implementation.

\subsubsection{Migration}
\label{sec:migration}

For each \idea-implementation pair $(C_j, \tilde{s}_{j,k})$, we prompt the model to generate multiple migrated implementations $\widetilde{M}_{j,k} = \{\tilde{m}_{j,k,1}, \tilde{m}_{j,k,2}, ..., \tilde{m}_{j,k,q}\}$ in target library $\mathcal{T}$
(\Cref{fig:data-gen}, Step 4). Each $\tilde{m}_{j,k,r}$ represents an attempt to migrate $\tilde{s}_{j,k}$ while preserving its public API.  This entails a multi-turn chat prompt that includes the original \idea $C_j$ and implementation $\tilde{s}_{j,k}$, and instructions asking the model to migrate the source implementation to use the target library. The prompt further instructs the model to keep the same structure and function signatures, and only update the internal logic and API calls to use the target library.
%
%As in the other steps we sample multiple migration attempts to increase the chances of success.

The output of this step is the set of raw migration triples 
$\widetilde{\mathcal{M}} = \bigcup_{j,k} \{\tilde{s}_{j,k}\} \times \widetilde{T}_{j,k} \times \widetilde{M}_{j,k}$, 
%$\widetilde{\mathcal{M}} = \bigcup_{j=1}^o \{ (\tilde{s}_{j,k}, \tilde{t}_{j,k,l}, \tilde{m}_{j,k,r}) : \tilde{s}_{j,k} \in \widetilde{S}_j, \tilde{t}_{j,k,l} \in \widetilde{T}_{j,k}, \tilde{m}_{j,k,r} \in \widetilde{M}_{j,k} \}$, 
where the union is taken over all \ideas $C_j \in \mathcal{C}$ and their implementations.
$\widetilde{\mathcal{M}}$ may include incorrect or only partially migrated examples, that are validated and filtered as described in Section~\ref{sec:validation}.

\subsection{Validation}
\label{sec:validation}

Not all sampled migration examples are valid. Initial or migrated implementations may be spurious or incomplete, tests may not run, etc.  
Thus, the validation step filters $\widetilde{\mathcal{M}}$ to produce a set of \emph{valid} migration triples $\mathcal{M} = \{(s_i, t_i, m_i)\}_{i=1}^m$. In a valid triple, both source and migrated implementations expose desired APIs, pass the same tests, and are exercised with sufficient test coverage. 

\vspace{1ex}
\noindent\textbf{Implementation Validation.}
First, we validate initial \idea implementations.
For each $(\tilde{s}_{j,k}, \tilde{t}_{j,k,\ell}) \in \widetilde{S}_j \times \widetilde{T}_{j,k}$, we execute tests $\widetilde{T}_{j,k}$ against $\widetilde{s}_{j,k}$, retaining pairs where all tests pass. 
%For each $\tilde{s}_{j,k}$, we execute all test suites in $\widetilde{T}_{j,k} = \{\tilde{t}_{j,k,1}, \tilde{t}_{j,k,2}, ..., \tilde{t}_{j,k,p}\}$ against $\widetilde{s}_{j,k}$, retaining all pairs $(\tilde{s}_{j,k}, \tilde{t}_{j,k,l})$ where all tests pass. 
Recall that we generate multiple potential test suites per implementation (\Cref{sec:testgen}); this step can therefore produce multiple pairs containing a given $s_{j,k}$. 
We run each test file $\tilde{t}_{j,k,\ell}$ independently using \texttt{pytest} inside a \texttt{docker} container that includes a pinned \py version and a large set of \py libraries, including $\mathcal{S}$ and its dependencies. 
%The output of this step are all the valid pairs of \texttt{(implementation, test)} pairs.

\vspace{1ex}
\noindent\textbf{Migration Validation.}  
For each valid $(s_{j,k}, t_{j,k,\ell})$ pair and  associated migration attempt $\tilde{m}_{j,k,r} \in \widetilde{M}_{j,k}$, we improve confidence in  equivalence by confirming that $\tilde{m}_{j,k,r}$ also passes tests $t_{j,k,\ell}$ (dropping those that do not).  
Because the migrated code is expected to preserve the original API, the same test suite is expected to apply (and those that do not are dropped). 

\vspace{1ex}
\noindent\textbf{Test-based Filtering.}
Finally, we also collect line-level coverage metrics for each run to build confidence that the test suites meaningfully exercise core logic. We simultaneously collect information on API usage for later analysis. We retain triples where shared tests achieve at least $60\%$ line coverage in both implementations.

To construct the final validated $\mathcal{M}$: for each implementation $\tilde{s}_{j,k}$ of \idea $C_j$ part of more than one valid triple, we select the one with the highest post-migration test coverage. We include implementations part of only a single valid triple directly. These selected triples become $(s_i, t_i, m_i) \in \mathcal{M}$. Thus, each validated implementation contributes at most one triple to $\mathcal{M}$, ensuring high-quality examples while maintaining \idea implementation diversity.

\section{Migration Script Synthesis}
\label{sec:synth}
\begin{figure*}[t]
    \centering
    \includegraphics[width=1\linewidth]{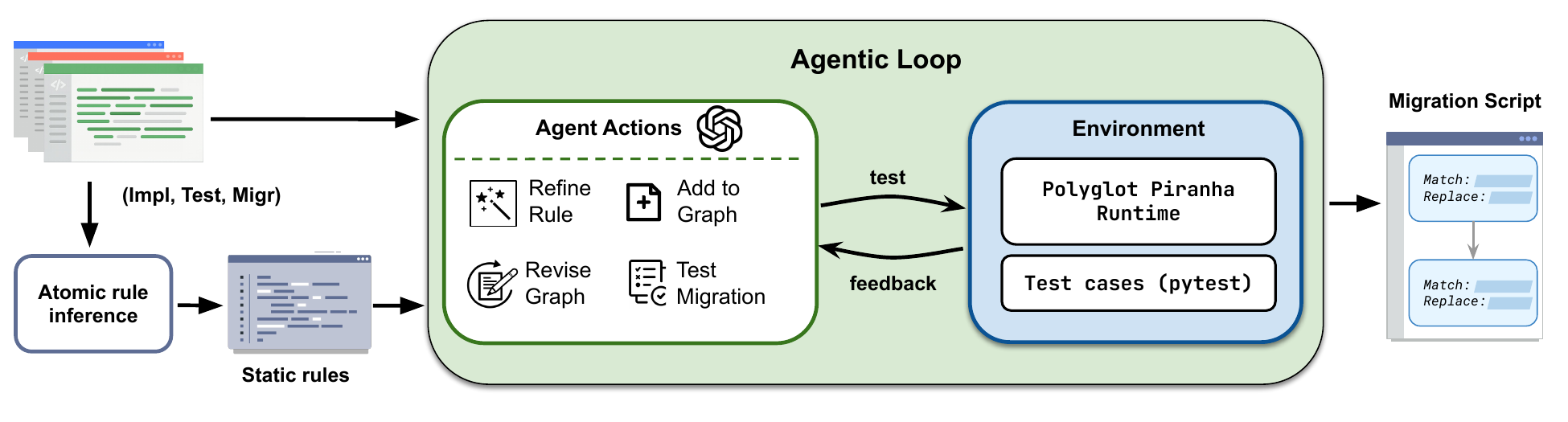}
    \caption{\small Agentic workflow of \tool's approach to migration script synthesis. Our pipeline leverages an atomic rule inference algorithm for creating an initial set of rules, which is then fed to an LLM agent for \piranha script generation. The agent iteratively applies changes to the script and gets feedback from the tests and \piranha's runtime to refine the rules. }
    \label{fig:agentic}
\end{figure*}

\tool converts migration triples $\mathcal{M} = \{(s_i, t_i, m_i)\}_{i=1}^m$ (Section~\ref{sec:data}) into executable rewrite scripts in \piranha, a state-of-the-art DSL and toolset for code transformation (Section~\ref{sec:piranha}). 
Our novel approach for script inference combines classical anti-unification with an agentic LLMs strategy. This hybrid is motivated by the structure of \piranha programs: local rewrite rules (graph nodes) connected by a high-level strategy (labeled edges). Classical anti-unification excels at inferring precise local rules from concrete examples, but cannot infer the orchestration strategy, scope, or cascading effects. We therefore complement the traditional synthesis with an LLM-based technique to organize mechanically-inferred atomic rules into a coherent transformation strategy. 

\looseness-1
For each migration triple $(s, t, m) \in \mathcal{M}$ script synthesis entails: (1)
\textbf{Rule Inference}: Automatically generate initial low-level transformation rules from source-target diff hunks (Section~\ref{sec:rule_inference}), and 
(2) \textbf{Agent-based Orchestration}: Use LLMs to refine and orchestrate the inferred rules into a rule graph (Section~\ref{sec:script_synthesis}).
This overall produces a migration program per successfully-processed triple in $\mathcal{M}$.

\subsection{\piranha}
\label{sec:piranha}

We infer migration scripts in \piranha, a domain-specific language for source-to-source transformation developed and deployed at Uber~\cite{piranha_pldi}. \piranha exemplifies a growing class of modern transformation DSLs~\cite{OpenRewrite, GritQL, astgrep} designed to support industrial-scale automated refactoring. These tools offer a structured, declarative alternative to ad hoc migration scripts.

To expand on key language constructs introduced in  \Cref{sec:motivation}: \piranha programs consist of local syntactic rewrite rules organized into a directed graph encoding application order, scoping constraints, and inter-rule dependencies. 
Each node defines a rule with a match and an optional rewrite clause; edges model control flow using scoped labels. I.e., an edge \( \mathcal{R}_1 \xrightarrow{\raisebox{-0.3ex}[0pt][0pt]{\tiny{Function}}} \mathcal{R}_2 \) reads as, ``{apply rule $\mathcal{R}_1$ and then apply rule $\mathcal{R}_2$ within the enclosing function where $\mathcal{R}_1$ was applied}''. 
Rules themselves are written as concrete syntax patterns (that closely resemble the target language’s surface syntax).
The concrete patterns are composed of strings and abstractions that match directly against concrete code. These abstractions are expressed via template variables (e.g., \lstinline|*:[x]*|), which bind to concrete sub-expressions. For example, the rule \lstinline|foo(*:[args+]*)| $\mapsto$ \lstinline|bar(*:[args]*)| rewrites \lstinline|foo(1,2,3)| as \lstinline|bar(1,2,3)|, where \lstinline|*:[args]*| binds to \lstinline|(1,2,3)| during the transformation.

\piranha programs run as depth-first traversal of the rule graph. The runtime begins with a queue of seed global rules and applies them depth-first until reaching a fixpoint. We refer interested readers to associated materials and documentation for full syntax and semantics~\cite{piranha_pldi,piranha_repo}.

Thus, for each migration triple $(s, t, m) \in \mathcal{M}$, we aim to synthesize a \piranha script $\mathcal{P} = (\mathcal{R}, \mathcal{E})$ where:
\begin{itemize}
\item $\mathcal{R} = \{r_1, r_2, ..., r_k\}$ is a set of transformation rules
\item $\mathcal{E} \subseteq \mathcal{R} \times \mathcal{L} \times \mathcal{R}$ is a set of directed edges with labels from $\mathcal{L} = \{\text{File}, \text{Function}, \text{Class}, ...\}$
\end{itemize}

Each rule $r_i \in \mathcal{R}$ is a tuple $(p_i, q_i)$ where $p_i$ is a match pattern, and $q_i$ is a replacement pattern. An edge $(r_i, \ell, r_j) \in \mathcal{E}$ indicates that after applying rule $r_i$, rule $r_j$ should be applied within scope $\ell$.

\subsection{Atomic Rule Inference}
\label{sec:rule_inference}

Let $\Delta(s, m)$ denote the set of diff hunks between source implementation $s$ and migration $m$. For each \emph{diff hunk} $h \in \Delta(s, m)$, let $h^- $ and $h^+$ denote the deleted and added lines, respectively.

We use an anti-unification algorithm to abstract these hunks into reusable rewrite rules. 
%Anti-unification computes the most specific generalization of two expressions---deriving a pattern that captures common structure while abstracting over differences using placeholders---making it ideal for synthesizing migration rules that preserve structural essence while enabling generalization. 
Anti-unification computes the most specific generalization of two expressions, producing a pattern that captures their common structure while abstracting over differences with placeholders. This makes it ideal for synthesizing migration rules that preserve the underlying structure yet allow for generalization.
We adapt an existing anti-unification algorithm $\mathcal{A}$ (from \melt~\cite{melt}) to infer an initial rule set:
$$\mathcal{R}_0 = \{r : r = \mathcal{A}(h^-, h^+) \text{ for } h \in \Delta(s, m) \text{ where } h^- \neq \emptyset\}.$$
Each rule abstracts shared elements in the \textit{before} and \textit{after} hunk.

For example, given the following hunk from a larger migration of \texttt{cryptography} to \texttt{pycryptodome}:

\begin{lstlisting}[language=diff, basicstyle=\tiny\ttfamily]
- encrypted_data = fernet.encrypt(data)
+ padded_data = pad(data, AES.block_size)
+ encrypted_data = iv + fernet.encrypt(padded_data)
\end{lstlisting}

Anti-unification, including heuristics to avoid overgeneralizing, produces: \lstinline|*:[x1]* = *:[x2]*.encrypt(*:[x3]*)|
\begin{quote}
$\mapsto$ \lstinline|padded_data = pad(*:[x3]*, AES.block_size)|\\
\phantom{$\mapsto$} \lstinline|*:[x1]* = iv + *:[x2]*.encrypt(padded_data)|
\end{quote}

%Transformation \textit{match-replace} rules are thus the unit element of programs written in \piranha's language (the nodes in the graph). Given a code diff representing changes between an \texttt{implementation, migration} pair, \tool uses \melt's anti-unification algorithm~\cite{melt} to automatically infer an initial set of match-replace rules for the \piranha script. Specifically, \melt attempts to generate a rule for each diff hunk between the source and target. Each rule abstracts shared elements in the \textit{before} and \textit{after} code.

%\tool processes this diff hunk using \melt's algorithm, which identifies and abstracts common subexpressions, resulting in the following generalized transformation rule:

Note that the algorithm we adapt does not support rule generation for hunks in which code is only added without deletion~\cite{melt}; we discuss this limitation in Section~\ref{sec:discussion}.

\subsection{Script synthesis}
\label{sec:script_synthesis}

Rule inference provides initial ruleset $\mathcal{R}_0$, but no composition strategy.
%Notice for example that the strategy in \Cref{fig:script-gen}, entails a dependency between \texttt{replace encrypt} and \texttt{replace decl}).  %
Agentic script synthesis constructs the full script $\mathcal{P} = (\mathcal{R}, \mathcal{E})$ through iterative refinement.
This refinement also allows \tool to correct initial rules $r$ that overgeneralize or are incorrect. 
An agentic approach for software tasks uses an LLM to iteratively take action, and observe its effects to refine a strategy for decomposing and solving a complex problem. 

\Cref{fig:agentic} illustrates the workflow.
It takes as input a migration triple $(s, t, m) \in \mathcal{M}$ and the initial rule set $\mathcal{R}_0$ inferred from diff $\Delta(s, m)$ between the source and target implementations. 
The model's task is to develop the overarching strategy comprising the full script while disambiguating conflicting rules and deciding on scopes and application order; and improving the transformation rules as necessary by, for example, resolving naming problems.  
The rest of this Section details key components of the approach. 

\vspace{1ex}
\noindent\textbf{System prompt.} We assume a model is unfamiliar with the relatively new \piranha language.
The system prompt explains language syntax, semantics, and runtime behavior, via: (1) examples of simple rules, explanations of concrete syntax, and guidance on matching/rewriting; (2) an explanation of rule graph, rule application strategies, scoping, and constructs for propagating template variables across rules; (3) examples of source code before and after migration, with detailed explanations of how  \piranha performs migration, and runtime traces, and finally (4) other guidelines, common pitfalls, and best practices. 

\vspace{1ex}
\noindent\textbf{Agentic Loop.}
For each migration triple $(s, t, m) \in \mathcal{M}$ and initial ruleset, \tool's agentic loop iteratively takes actions and observes their effects until it produces a \piranha program that transforms the source $s_i$ into a migration $\hat{m}_i$ that passes the original test suite $t_i$, or until it completes 10 iterations.

The first iteration takes a partially-complete task template containing the migration triple and initial \tool-inferred rules.
We prompt the model to generate a \piranha program that transforms $s_i$ into $m_i$, following system guidelines. The prompt specifies that the produced program need not produce an exact token-level match with $m_i$, but should capture the intended semantic transformation. We (and the agent) verify correctness by running code produced by the eventual program on test suite $t_i$. 

Each iteration, the model selects an action and executes it in a controlled environment, with result returned as feedback: 

\begin{itemize}
\item \textbf{Refine / Create an Atomic Rule.}
The model can create a new match-replace rule, or refine an existing one, and apply it to $s_i$ in isolation to observe an atomic rule's behavior independently of other transformations. %\tool executes the new rule in the controlled environment using the \piranha interpreter.
%and provides the result as feedback to the next iteration

\item\textbf{Add a Rule to the Graph.}
The model can add one or more new rules to the (initially empty). The action specifies rule order and scope. 
%\tool updates the new graph to incorporate new rules and connections and applies it to $s_i$.
%providing the result as feedback.  

\item\textbf{Revise the Current Rule Graph.}
A model can revise the graph beyond atomic rule refinement by modifying portions of the graph, or regenerating it, enabling high-level restructuring of the transformation strategy. % \tool applies the revised graph to $s_i$.  %providing the result as feedback. 

\item\textbf{Test Migration.}
The model can use tests $t_i$ to validate the current rule graph by applying it to $s_i$ and running tests $t_i$ against the result.   If the tests pass, the loop terminates. 
\end{itemize}

\noindent\textbf{Environment.}
\tool executes actions within a structured environment and provides results back to the model to guide the next agentic step. For rule refinement and integration, and rule graph revision, \tool applies generated $P$ to $s_i$ using the \piranha interpreter and returns the resulting migrated program as feedback.  \tool provides additional feedback when available, e.g., if applying $P$ fails due to a syntax or \piranha runtime error, \tool returns the error message with a set of likely causes and correction suggestions. These are derived from common failure modes and known pitfalls in the language engine, akin to compiler diagnostic messages. 
%
%If the script executes but does not change $s_i$, \tool provides a set of potential reasons, again with common pitfalls that we manually wrote from our experience with the tool.
If the script runs but does not modify $s_i$, \tool offers possible explanations and common pitfalls, based on our experience with the tool.
For testing, \tool applies the script and runs the transformed code inside 
a containerized test environment. If all tests pass, \tool records the migration as successful; Otherwise, it returns the test failures and error messages for the next iteration. 

Once we have the migration scripts, we can use them to update any codebase from the source to the target library, as the rules are project-agnostic. The developer responsible for the migration can review these concise rules to understand the necessary changes and then use \piranha to automatically apply them.

\section{Evaluation}
\label{sec:eval}

We answer four research questions:

\begin{enumerate}[label=\textbf{{RQ}\arabic{*}.}, leftmargin=*, widest=10]
    \item \textbf{(End-to-end-effectiveness): How effectively does \tool generate \piranha migration scripts?} 
    We evaluate the extent to which the entire pipeline produces working \piranha scripts. 

    %We evaluate the synthesis pipeline by measuring (a) the percentage of migration examples for which \tool synthesizes a \piranha script that, when applied to the example, successfully performs the migration and passes all associated tests, and (2) how well the synthesized scripts generalize to other implementations of the initial \idea. 

    \item \textbf{(Real-world applicability): How useful are \tool's scripts for migrating real-world client code?}
    We test whether the \tool-produced \piranha scripts, learned from synthetic examples, successfully migrate real open-source projects, to evaluate practical utility.

    %We apply the inferred scripts to open-source projects and measure how many rewrites they perform and if the project's tests still pass after migration.

    \item \textbf{(Comparison with prior work): How does \tool compare to \melt, a prior tool for automated refactoring?}
    We compare \tool to \melt, the most-closely-comparable prior approach for migration rule inference in Python.

    %We compare \tool to \melt, the most-closely-comparable prior approach for migration rule inference.  \melt infers migrations to update client code with respect to breaking library changes; we evaluate and compare it on the synthetic examples generated by \tool.
    %While \melt targets breaking changes, it is the most comparable prior work in migration rule inference. We evaluate \melt on synthetic examples generated by \tool, measure the number of examples for which it can infer migration rules, and compare this to the performance of \tool.

    \item \textbf{(Data quality): What is the quality of the synthetic migration examples from the data generation pipeline?} 
    A core novelty of \tool is its bootstrapping via synthetic examples.  We assess the extent to which these examples are valid and diverse, and provide sufficient coverage of both source and target APIs and implementations. 
    %We measure how often the pipeline can generate \ideas with  implementations in the source and target libraries that pass the generated tests, as well as the test coverage and API diversity of the migration examples.\clg{I feel like this one could be slightly better motivated.  I agree it belongs, it just doesn't do a great job justifying itself.}
\end{enumerate}

\subsection{Experimental Setup}

\noindent\textbf{Implementation.}
We implement \tool in \py. 
Both synthetic data generation and agentic synthesis are model-agnostic and use the OpenAI Chat Completion API~\cite{openai-python} (adopted by most major LLM API providers).
%For data generation, we use \texttt{gpt-4o-mini} (a small model) due to its cost effectiveness, drawing on prior results suggesting that more attempts offset the limitations of smaller models~\cite{snell2024scaling}. 
For data generation, we use the cost-effective \texttt{gpt-4o-mini} (a small model), relying on prior work showing that more generation attempts can compensate for smaller models~\cite{snell2024scaling}.
For synthesis, we use \texttt{gpt-4.1} \cite{openai-gpt4-1}, a current state-of-the-art model with more advanced capabilities and larger context window (1M tokens), required for our large system prompt and chained interactions.

\vspace{1ex}
\noindent\textbf{Migration tasks.} 
We evaluate on 10 migration tasks across 20 \py libraries; the left-hand-side of \Cref{tab:overall_results} summarizes.  These tasks were selected to span a diverse range of popular libraries and their alternatives across multiple domains, which prior work has also targeted~\cite{pyevolve, melt, soar, islam2025using}.
%including CLI construction, data science, cryptography, templating engines, date and time utilities, HTTP requests, and logging. These domains have also
We focus on \py because there is a well-documented lack of refactoring tools for \py~\cite{api_evolution_review}, despite the fact that it is among the most popular programming languages~\cite{tiobe}; this also makes it a good target for LLM-based tooling~\cite{jiang2024survey}.

\vspace{1ex}
\noindent\textbf{Settings.}
We generate 100 \ideas per migration task $(\mathcal{S},\mathcal{T})$.  We generated 5 source implementations $\tilde{s}_{j,k}$ for each \idea $C_j \in \mathcal{C}$; 5 test files $\tilde{t}_{j,k,\ell}$ per source implementation $\tilde{s}_{j,k}$; and 5 migration attempts $\tilde{m}_{j,k,\ell}$ per source implementation  $\tilde{s}_{j,k}$.
This initially yields 500 implementations, 2,500 test files, and 2,500 migration attempts per task; these are validated and filtered as described in Section~\ref{sec:validation}. 
%Validation produces a set of migration triples $\mathcal{M} \subseteq S \times T \times M$ where each $(s, t, m) \in \mathcal{M}$ satisfies the filtering criteria: $\text{coverage}(s, t) \geq 0.6 \land \text{coverage}(m, t) \geq 0.6$.
%
We run our synthesis pipeline on each validated triple to attempt to generate a \piranha script $\mathcal{P} = (\mathcal{R}, \mathcal{E})$. We allow the agent up to 10 iterations to produce the script. 

We chose these numbers (10, 5) to balance cost and coverage: based on model pricing, this setup allowed us to stay within a \$100 total budget while still producing meaningful results.
We generated more migration attempts because migration is a harder problem; multiple attempts increase the chance of success~\cite{snell2024scaling}.

A synthesis attempt is considered \emph{successful} if the script replaces all source library API calls  ($\text{APIs}(\mathcal{P}(s), \mathcal{S}) = \emptyset$) and, if when applied to $s_j$, the migrated code passes the same tests as $m_{j,k}$ ($\mathcal{P}(s) \equiv_t m$). 

\begin{table*}
\caption{\small Overview of migration tasks with synthesis results. Valid Triples: generated migration triples with >60\% coverage. Success: percentage of migration triples for which each tool successfully generated a validated \piranha script. 
 Sibling Success: sibling implementations that at least one synthesized \piranha script migrates successfully out of the total number of potentially-migratable sibling implementations. %\ruben{It was unclear how sibling implementation relates to the valid triples. What is the upper bound on the number of siblings? The valid source implementations that were generated?}
 }
\label{tab:overall_results}
\small
\resizebox{\textwidth}{!}{%
\begin{tabular}{lllc|rrrrr}
\toprule
\multicolumn{4}{c}{\textbf{Migration Task}} & \textbf{Valid} & \multicolumn{2}{c}{\textbf{Success (\%)}} & \textbf{Sibling} \\
\textbf{Source} & \textbf{Target} & \textbf{Domain} & \textbf{Stars (S / T)} & \textbf{Triples} & \textbf{\tool} & \textbf{MELT} & \textbf{Success} \\
\midrule
\href{https://docs.python.org/3/library/argparse.html}{\texttt{argparse}} & \href{https://github.com/pallets/click}{\texttt{click}} & CLI & N/A / 16.4k & 215 & 44.2 & 17.2 & 91/170 \\
\href{https://github.com/pallets/jinja}{\texttt{jinja2}} & \href{https://github.com/sqlalchemy/mako}{\texttt{mako}}  & Templating engines & 10.9k / 389 & 15 & 13.3 & 0.0 & 0/3 \\
\href{https://docs.python.org/3/library/json.html}{\texttt{json}} & \href{https://github.com/ijl/orjson}{\texttt{orjson}} & JSON Serialization & N/A / 6.9k & 269 & 96.7 & 57.6 & 258/291 \\
\href{https://docs.python.org/3/library/logging.html}{\texttt{logging}} & \href{https://github.com/Delgan/loguru}{\texttt{loguru}}  & Logging & N/A / 21.7k & 114 & 85.1 & 72.8 & 64/93 \\
\href{https://github.com/lxml/lxml}{\texttt{lxml}} & \href{https://www.crummy.com/software/BeautifulSoup/}{\texttt{BeautifulSoup}} & HTML/XML parsing & 2.8k / N/A & 32 & 59.4 & 0.0 & 0/21 \\
\href{https://github.com/pandas-dev/pandas}{\texttt{pandas}} & \href{https://github.com/pola-rs/polars}{\texttt{polars}}   & Data analysis  & 45.5k / 33.8k & 0 & - & - & - \\
{\texttt{pathlib}} & \href{https://github.com/PyFilesystem/pyfilesystem2}{\texttt{pyfilesystem2}} & Filesystem abstraction & N/A / 2.0k & 21 & 52.4 & 28.6 & 1/15 \\
\href{https://github.com/pyca/cryptography}{\texttt{cryptography}} & \href{https://github.com/dlitz/pycrypto}{\texttt{pycryptodome}} & Cryptography & 7.0k / 3.0k & 79 & 48.1 & 6.3 & 3/46 \\
\href{https://github.com/psf/requests}{\texttt{requests}} & \href{https://github.com/encode/httpx}{\texttt{httpx}} & HTTP clients (async, sync) & 52.9k / 14.1k & 65 & 76.9 & 12.3 & 39/63 \\
\href{https://docs.python.org/3/library/time.html}{\texttt{time}} & \href{https://github.com/sdispater/pendulum}{\texttt{pendulum}} & Date/time & N/A / 6.4k & 60 & 78.3 & 11.7 & 35/73 \\
%\midrule
%\textbf{Average} & & & & \textbf{96.7} & \textbf{61.6} & \textbf{22.9} & \textbf{54.6} \\
\bottomrule
\end{tabular}
} % resizebox
\end{table*}

\subsection{RQ1: End-to-end effectiveness}
\label{sec:rq-end-to-end}

\noindent\textbf{Methodology.}
We first evaluate our synthesis pipeline end-to-end. We report, for each migration task, the number of valid triples generated over all \ideas (\Cref{tab:overall_results}, 5th column, ``Valid triples''); this is the output of data generation (Section~\ref{sec:data}).  
Success rate (for \tool, \Cref{tab:overall_results} column 6) reports the percentage of valid examples on which script synthesis produce a successful \piranha program. We investigate synthetic data quality and diversity in Section~\ref{sec:rq-data-quality}.
% table 2: /Users/danielramos/Desktop/Dev.nosync/piranha-synthesis/experiment_results/generate_success_table.py
%
% script for rq1: /Users/danielramos/Desktop/Dev.nosync/piranha-synthesis/src/rq1/find_never_successfully_migrated.py
%
% tables3, 4: /Users/danielramos/Desktop/Dev.nosync/piranha-synthesis/src/analysis/generate_table.py
%
% table 5 was manual, but loc counted with : /Users/danielramos/Desktop/Dev.nosync/piranha-synthesis/rq4_repos/count_loc.py
%
To evaluate script generalizability, we test each $\mathcal{P}_{j,k}$ (from triple $(s_{j,k}, t_{j,k,\ell}, m_{j,k,r})$) on \emph{sibling implementations} $\{s_{j,i} : i \neq k\}$ of the same \idea $C_j$. We report how many siblings are successfully migrated by the generated scripts, per the tests.

\vspace{1ex}
\noindent\textbf{Results.}
\Cref{tab:overall_results} shows results. \tool successfully generated valid migration triples for nine of ten tasks, producing an average of 87 filtered, valid triples per task (the \texttt{pandas → polars} exception is discussed in \Cref{sec:rq-data-quality}).  
Using these validated examples, \tool's migration synthesis script succeeded, on average, $61.6\%$ of the time, using a single trial. This rate could likely improve further with test-time scaling~\cite{snell2024scaling}; we leave this to future work, given cost. 

The last column of \Cref{tab:overall_results} shows that the scripts migrated 63.3\% of sibling implementations (491/774, last column of \Cref{tab:overall_results}).  
Performance was particularly strong for simple, one-to-one API replacements like 
\texttt{\footnotesize json → orjson} (88.6 \%) while migrations involving embedded DSLs (\texttt{jinja2 → mako}, \texttt{lxml → beautifulsoup}) generalize less well. 

These libraries support dynamic webpage generation and embed templating languages within \py string literals, which the \piranha engine cannot transform; this is a limitation of the \piranha language itself~\cite{piranha_pldi}.  Despite this constraint, our synthesized rules effectively capture common usage patterns across most migration tasks.

\subsection{RQ2: Real-world applicability}
\label{sec:rq-real-world}
\begin{table*}
\caption{\small
Inferred migration scripts applied to real-world repositories. Each row represents the application of a migration script to a repository that uses the source library. \textbf{Stars} and \textbf{KLoC} refer to the GitHub popularity and codebase size. \textbf{Rewrites} is the number of times a \piranha rule triggered in the project. Passing Tests (Before/After/\%): Number of passing test cases pre- and post-migration, with the percentage of tests preserved (post/pre), indicating basic functionality is maintained.
%\textbf{Passing Tests (Before \& After)}: Test cases pass both before and after migration, a basic signal of functionality preservation.
}
\label{tab:real_world}
\small
\resizebox{\textwidth}{!}{%
\begin{tabular}{llrr|rrrr}
\toprule
 &  & & &  & \multicolumn{3}{c}{\textbf{Passing Tests}} \\
\textbf{Migration Task} & \textbf{Repository}&  \textbf{Stars} &  \textbf{KLoC} & \textbf{Rewrites} & \textbf{Before} & \textbf{After} & \textbf{\%}\\
\midrule
\multirow{3}{*}{\texttt{json → orjson}}
    & \texttt{BIM2SIM/bim2sim} & 57 & 33.8 & 46 & 173 & 149 & 86\%\\
    & \texttt{kaapana/kaapana} & 196 & 69.8 & 317 & 45 & 43 & 96\%\\
    & \texttt{SoCo/SoCo} & 1500 & 17.4 & 7 & 233 & 233 & 100\%\\
\midrule
\multirow{3}{*}{\texttt{logging → loguru}}
    & \texttt{fastqe/fastqe} & 172 & 0.7 & 16 & 2 & 2 & 100\%\\
    & \texttt{mie-lab/trackintel} & 229 & 11.2 & 14 & 411 & 411 & 100\%\\
    & \texttt{pywbem/pywbem} & 42 & 171.7 & 16 & 2499 & 0 & 0\%\\
\midrule
\multirow{3}{*}{\texttt{lxml → BeautifulSoup}}
    & \texttt{acl-org/acl-anthology} & 528 & 19.4 & 111 & 433 & 433 & 100\%\\
    & \texttt{pyxnat/pyxnat} & 73 & 7.7 & 8 & 62 & 62 & 100\%\\
    & \texttt{w3af/w3af} & 43 & 201.8 & 10 & 2547 & 2547 & 100\%\\
\midrule
\multirow{3}{*}{\texttt{cryptography → pycryptodome}}
    & \texttt{aomail-ai/aomail-app} & 134 & 2.0 & 1 & 13 & 11 & 85\%\\
    & \texttt{BM/qpylib} & 32 & 2.0 & 4 & 139 & 127 & 91\%\\
    & \texttt{DMcP89/harambot} & 32 & 1.6 & 8 & 8 & 0 & 0\%\\
\midrule
\multirow{3}{*}{\texttt{requests → httpx}}
    & \texttt{Kildrese/scholarBibTex} & 33 & 0.1 & 5 & 1 & 1 & 100\%\\
    & \texttt{wjohnson/pyapacheatlas} & 175 & 13.9 & 22 & 128 & 128 & 100\%\\
    & \texttt{databricks/databricks-sdk-py} & 433 & 108.6 & 23 & 157 & 157 & 100\%\\
\midrule
\multirow{3}{*}{\texttt{time → pendulum}}
    & \texttt{clld/clld} & 57 & 9.1 & 4 & 308 & 308 & 100\%\\
    & \texttt{fbpic/fbpic} & 192 & 25.5 & 43 & 6 & 5 & 83\%\\
    & \texttt{qutip/qutip} & 1800 & 41.2 & 29 & 1585 & 147 & 9\%\\
\bottomrule
\end{tabular} 
} % resizebox
\end{table*}

\looseness-1
\noindent\textbf{Methodology.}
To evaluate whether our inferred migration scripts generalize beyond synthetic examples, we applied them to 18 Python projects hosted on GitHub. We used \texttt{Sourcegraph}~\cite{sourcegraph} to collect a convenience sample of codebases containing source API calls that were present in our generated scripts, sorting them by start, and prioritizing codebases with test suites. 
Due to the high engineering effort required to make each project build and test successfully, we conducted this evaluation for only six of the ten migration tasks.

We manually reviewed each project to identify the most-relevant generated migration script that best matched observed API usage patterns. We containerized execution using \texttt{Dockerfile}s based on each project's documented setup, ran tests to establish baseline functionality, then applied the selected \piranha script and re-ran tests. We updated \texttt{requirements.txt} to include target dependencies, but retained source dependencies, because our scripts do not usually cover all API usages within a library.  We did not measure transformation time, as \piranha's engine executes near-instantaneously.
These non-trivial codebases typically span hundreds of files and thousands of lines of code.

\vspace{1ex}
\noindent\textbf{Results.}
\Cref{tab:real_world} reports the number of rewrites (i.e., the number of times a particular \piranha rule was triggered to transform code) per project, as well as the number of passing tests before and after migration. By checking whether pre-existing tests still pass, we validate that the functionality maintainers deemed important is preserved.
In many cases, all tests passed post-migration, despite scripts being inferred from synthetic examples.
%In many cases, all tests continued to pass post-migration, despite the scripts being inferred from synthetic examples.
For example, for the \texttt{logging → loguru} migration, the inferred script preserved test behavior in both the \texttt{trackintel} and \texttt{fastqe} projects and triggered a substantial number of rewrites. Similarly, \tool's scripts for \texttt{json → orjson} and \texttt{requests → httpx} applied dozens to hundreds of edits across real-world projects while preserving functional behavior.
Upon manual inspection, we found that test failures were typically due to incomplete migrations: the scripts handled part of the migration correctly but required further refinement or additional transformations. 

\subsection{RQ3: Comparison with prior work}
\label{sec:rq-compare}

\begin{figure}
    \centering
    \includegraphics[width=\columnwidth]{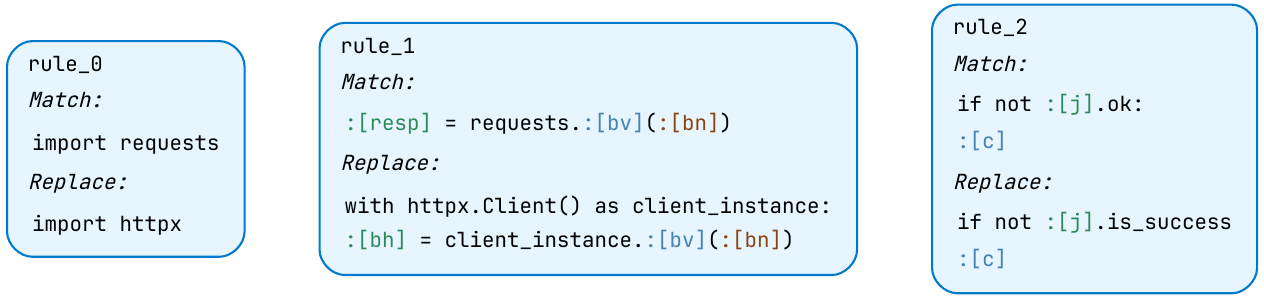}\vspace{0.2em}\hrule\vspace{0.2em}
\includegraphics[width=\columnwidth]{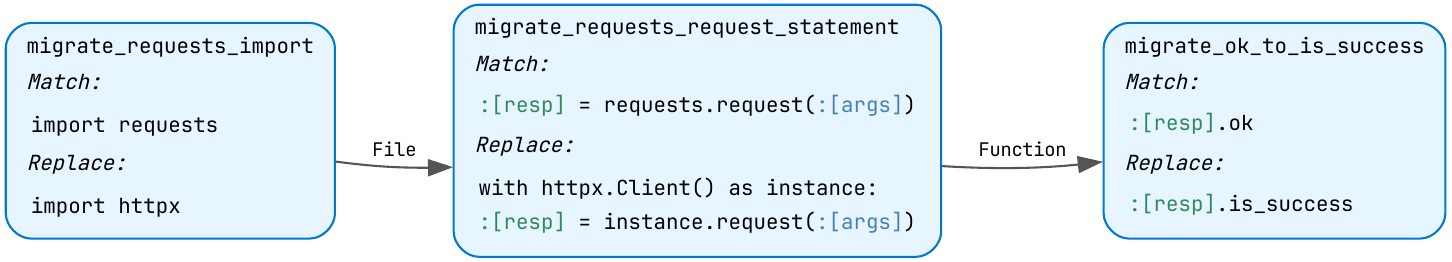}
    \caption{\small A program generated by \melt (top) and \tool (bottom) from a migration example for \texttt{\footnotesize requests → httpx}. \melt's rules contain generic placeholder names and overabstracts; \tool's script includes scoped application and semantically meaningful names. } %\ruben{This figure is too small. It was very hard to read on paper; there is a lot of white space around the figure. The text should be larger, and we can reduce the white space.}}
    \label{fig:rq3-results}
    \vspace{-3mm}
\end{figure}

\noindent\textbf{Methodology.}
We compare our synthesis approach with \melt~\cite{melt}, a state-of-the-art tool for automated \py refactoring.
%\melt because it is one of the few tools that target \py and builds on well-established transformation DSL (\texttt{comby}~\cite{comby_pldi, comby_website}, see \Cref{sec:related}).  
%\melt infers synthesis scripts from migration pairs mined from developer pull requests on updated libraries, rather than generated or synthetic examples as \tool does. To enable a fair comparison (not all of our migrations are associated with such pull requests) that more effectively isolates the hybrid synthesis in \tool, we run \melt on the generated migration examples. 
\melt is one of the few tools designed for Python and is built on the transformation DSL \texttt{comby}~\cite{comby_pldi, comby_website} (see \Cref{sec:related}).
While \tool uses generated or synthetic examples, \melt infers migration scripts from real migration pairs mined from developer pull requests. For a fair comparison (not all of our migrations are associated with such pull requests) that more effectively isolates the hybrid synthesis in \tool, we run \melt on the generated migration examples.

\vspace{1ex}
\noindent\textbf{Results.}
\Cref{tab:overall_results}, columns 6 and 7, compares the two approaches.   \tool outperforms \melt on all tasks for which valid triples are available, often by a large margin. It achieves a 61.6\% (average) success rate compared to \melt's 22.9\%, despite both tools starting from the same validated examples.
The fact that \melt can extract transformation rule from many of the synthetic examples provides additional evidence of their quality, and suggests the data generation pipeline may enhance other migration inference tools.  However, \melt's rules are created in isolation.  
\tool's agentic pipeline, by contrast, orchestrates these kinds of rules into coherent transformation strategies, effectively taking advantage of the expressive power of the \piranha language. \Cref{fig:rq3-results} illustrates this difference: \tool synthesizes interconnected rules with explicit scoping (per-function, per-file) that enable cascading transformations without overgeneralization; \melt produces disconnected global rules with generic placeholders, which can prove especially problematic on large codebases~\cite{piranha_pldi}.

These results validate that (1) the generated synthetic examples provide a useful independent signal for migration inference, (2) the agentic approach to rule composition meaningfully improves over the current state-of-the-art, with intelligent orchestration of inferred rules importantly contributing to migration synthesis.

%\begin{highlight}
%    \textsc{Spell}'s agentic approach consistently outperforms \melt across all migration tasks, while \melt itself benefits from the synthetic examples generated by \tool.
%\end{highlight}

\subsection{RQ4: Data Quality}
\label{sec:rq-data-quality}

\begin{table*}
\caption{\small Quality of generated migration examples: \idea success rates, API diversity, and test coverage}
\label{tab:data_quality}
\small
\resizebox{\textwidth}{!}{%
\begin{tabular}{lrrrrrrr}
\toprule
\textbf{Migration Task} & \multicolumn{2}{c}{\textbf{\ideas (out of 100)}} & \multicolumn{2}{c}{\textbf{Distinct APIs}} & \multicolumn{2}{c}{\textbf{Avg. Coverage}} & \textbf{Validated} \\
\cmidrule(lr){2-3} \cmidrule(lr){4-5} \cmidrule(lr){6-7}
& \textbf{Implemented} & \textbf{Migrated} & \textbf{Source} & \textbf{Target} & \textbf{Source} & \textbf{Target} & \textbf{Examples} \\
\midrule
\texttt{argparse → click} & 74 & 72 & 11 & 16 & 0.89 & 0.93 & 215 \\
\texttt{jinja2 → mako} & 24 & 14 & 9 & 4 & 0.86 & 0.86 & 15 \\
\texttt{json → orjson} & 96 & 96 & 4 & 3 & 0.70 & 0.70 & 269 \\
\texttt{logging → loguru} & 58 & 52 & 12 & 6 & 0.79 & 0.79 & 114 \\
\texttt{lxml → beautifulsoup} & 59 & 23 & 11 & 9 & 0.75 & 0.75 & 32 \\
\texttt{pandas → polars} & 58 & 2 & 5 & 4 & 0.43 & 0.43 & 0 \\
\texttt{pathlib → pyfilesystem2} & 64 & 23 & 10 & 17 & 0.71 & 0.70 & 21 \\
\texttt{cryptography → pycryptodome} & 72 & 52 & 59 & 25 & 0.73 & 0.72 & 79 \\
\texttt{requests → httpx} & 64 & 60 & 8 & 10 & 0.66 & 0.64 & 65 \\
\texttt{time → pendulum} & 81 & 55 & 9 & 34 & 0.61 & 0.61 & 60 \\
\midrule
\textbf{Average} & \textbf{65.0} & \textbf{44.9} & \textbf{13.8} & \textbf{12.8} & \textbf{0.71} & \textbf{0.71} & \textbf{87.0} \\
\bottomrule
\end{tabular}
} % resizebox
\end{table*}

\noindent\textbf{Methodology}
We characterize the quality of the synthetically generated data across several dimensions: 
\begin{enumerate}
    \item  \textbf{Use case implementation:} the number of the 100 generated \ideas  $C_j \in \mathcal{C}$ that yields at least one implementation $s_{j,k}$ that passes at least one generated test harness.   
    \item \textbf{Use case migration:} the number of implemented \ideas that are migrated to the target library (how many \ideas have at least one $s_i$ that is associated with a success $m_i$). 
    \item \textbf{API diversity:} number of distinct API methods used from both $\mathcal{S}$ and $\mathcal{T}$.\footnote{Due to \texttt{\scriptsize Python}’s dynamic typing, this heuristic should be considered a lower bound.}
    %\footnote{Note that this is heuristic (due to \texttt{\scriptsize Python} ’s dynamic typing), so it should be considered a lower bound.} 
    %
    We compute this using the Jedi static analyzer~\cite{jedi} on validated triples $(s, t, m) \in \mathcal{M}$.
    \item \textbf{Test coverage:} via line-level coverage of the generated tests on both source and target implementations (coverage must be at least $0.6$, per our definition of validity).
    \item \textbf{Validated examples:} The total number of validated examples from all generated examples, post-filtering; this determines the training data available for script synthesis. 
\end{enumerate}

Together, these metrics characterizes the quality of the generated synthetic data, in terms of how well they cover or represent a migration task, and the degree to which they are adequately tested.

\vspace{1ex}
\noindent\textbf{Results.}
\Cref{tab:data_quality} shows results across all data quality metrics. 
\tool successfully implemented the majority of generated \ideas, with an average of 65 out of 100 \ideas yielding at least one working implementation. Success rates ranged from 24 (\texttt{jinja2 → mako}) to 96 (\texttt{json → orjson}). Upon manual inspection of \texttt{jinja2 → mako}, we found that most test code for this task used string comparisons for expected output, which the model struggled to get right. These tests were also not transferable between libraries, as each uses a different string representation to denote placeholders in the templates.

From the pool of \ideas implemented at least once, \tool successfully migrated an average of 44.9 \ideas while maintaining functional equivalence per the generated tests. The outlier is \texttt{pandas → polars} with only 2 successful migrations despite 58 successful implementations. Manual inspection revealed that generated test harnesses relied on pandas-specific constructs (e.g., \texttt{DataFrames}) that were incompatible with \texttt{polars}.

Our analysis of API diversity using Jedi identified an average of 13.8 distinct source library APIs and 12.8 target library APIs per migration task. These likely correspond to the most frequently used APIs that the underlying LLM encountered during pre-training~\cite{jiang2024survey}. The \texttt{cryptography → pycryptodome} task showed the highest diversity (59 source, 25 target APIs), while simple tasks like json → orjson used fewer APIs (4 source, 3 target). %\clg{something more to say here? Does it look like diversity scales with complexity, at least anecdotally?}

The generated tests achieved an average coverage of 71\% for both source and target implementations, well above our 60\% filtering threshold. Coverage ranged from 61\% (\texttt{time → pendulum}) to 89\%/93\% (\texttt{argparse → click}).  This suggests that tests meaningfully exercise implementation logic, while leaving room to improve; LLMs are known to struggle with unit test generation~\cite{catlm}. %\clg{Any reason we think it did better or worse on one versus the other?}

After applying all filtering criteria, \tool produced an average of 87 validated migration triples per task (excluding \texttt{pandas → polars}). 
Overall, these examples are testable and diverse (covering an average of 13–14 distinct APIs per library), but also exhibit strong coverage (71\% on average) on both source and migrated code.
This supports their use as input for downstream synthesis. %, and shows that even small models can surface structured, high-coverage migration examples when prompted systematically.

\section{Discussion and Limitations}
\label{sec:discussion}

\noindent\textbf{Quality and Representativeness of Synthetic Data.}
Our pipeline relies on synthetic examples generated by a comparatively small model (\texttt{gpt4o-mini}) to keep the cost of data generation manageable. We show that this is sufficient for both our approach, and that the generated data is usable by a previous migration inference approach. 
However, programs generated by LLMs are typically biased towards especially well-used parts of each API. Corner-case behaviours, error-handling paths, and rarely used configuration parameters are likely under-represented. As larger models trained on more data become cheaper, we expect both the breadth of API usage and correctness and quality of the generated programs to improve. A complementary direction is to augment the \idea–implementation prompts with retrieval over real code repositories.  This could be especially helpful for a use case enabling a client to articulate relevant API methods to guide sample generation, which could improve coverage of migration rules on real-world repositories. 

\vspace{1ex}
\noindent\textbf{Applicability.}
We represent our migration in the \piranha language because of its expressive and concise rule-graph abstraction, as well as high performance. However, \piranha still has limitations, in particular, it cannot yet rewrite code inside string literals, formatted templates, or dynamically generated constructs. This limitation was the main reason for the low success rate on the \texttt{jinja2$\rightarrow$mako} task. Integrating multi-language parsing in \piranha, or switching to a hybrid transformation engine that supports embedded DSLs could improve performance on this task.

\Cref{sec:eval} shows that the technique is promising for some migration tasks (e.g., when the code manipulates standard data types or produces well-defined outputs that are easy to compare even across libraries) and inadequate in others. A main limitation in real-world usage is the many variations of code found in open-source projects. 
The atomic rules inferred by \melt capture concrete diff hunks faithfully but can be either too specific, failing to match slight syntactic variations, or too general, introducing false positives. \tool's agentic loop alleviates some of these issues by refining and testing rules, yet it remains fundamentally example-bound: if a usage pattern does not appear in any migration triple, the resulting script will not handle it. \texttt{Pycraft}~\cite{dilhara2024unprecedented}'s technique on rule generalization using LLMs could be integrated into our workflow to generate further equivalent script variations for a broader application. 
Future work can also complete partial migrations by incorporating project-specific API usage patterns to refine migration scripts and by using failing test feedback to fix remaining issues.

\vspace{1ex}
\noindent\textbf{Tests.}
A core pillar of \tool's validation process is the use of automatically generated tests to assess the correctness of both implementations and their migrated counterparts. 
While this enables large-scale validation without manual effort, it introduces some limitations.
First, the quality of the tests is inherently tied to the generative model's ability to produce robust and meaningful test harnesses. Although we mitigate this by sampling multiple test files per implementation and applying a filter based on coverage as explained in \Cref{sec:testgen}, the tests may still be shallow or fail to exercise edge-case behavior. 
%
%In particular, the tests are designed to confirm functional equivalence at a coarse granularity (e.g., returning the same output for a given input) but may miss subtle behavioral divergences or even side effects.
%
In particular, tests confirm only basic functional equivalence (e.g., matching outputs), so they may miss subtle behavioral divergences or side effects.
Second, our approach assumes that passing the same test suite is a sufficient proxy for semantic equivalence. While this assumption is reasonable for many cases, it is not formally guaranteed. The test suites may lack completeness or specificity.
Third, our coverage-based filtering helps ensure that the tests exercise a non-trivial portion of the implementation logic, but it does not guarantee semantic soundness. For instance, tests that exercise a large portion of the code but contain weak assertions (e.g., asserting only that no exceptions are raised) may still pass despite incorrect behavior. Other validation techniques (e.g., mutation testing~\cite{mutation}) could help mitigate this~issue. %\ruben{Should we talk about the test cases of the real-world projects? Since our approach worked on test cases that were not generated by us or by LLMs, it increases the assurance on our approach, even if the generated tests have some flaws.}

\section{Related Work}
\label{sec:related}

\noindent\textbf{Code Transformation Toolsets.}
Code transformation toolsets vary widely. Imperative frameworks for AST rewriting include Java’s \texttt{errorprone}\cite{errorprone} and \texttt{OpenRewrite}\cite{OpenRewrite}. \texttt{clang libtooling}\cite{clang} provides declarative matching via AST matchers, along with an imperative framework for rewriting code. 

Purely declarative tools for code transformation also exist, for example, \texttt{coccinelle}~\cite{coccinelle}, \texttt{GritQL}~\cite{GritQL}, \texttt{ast-grep}~\cite{astgrep}, \texttt{comby}~\cite{comby_pldi},  and \piranha~\cite{piranha_pldi}.
\texttt{comby} is the intellectual predecessor of \piranha, which we use in this paper. \piranha builds on the \texttt{comby} model by allowing rules to be composed for fine-grained and scoped rewrites using cascading transformations (edges), explicit scopes, and filters. We chose to represent our migration scripts in \piranha because of its demonstrated success in industrial-scale applications~\cite{piranha_pldi} and its syntactic simplicity, which makes is amenable for synthesis.

\vspace{1ex}
\noindent\textbf{Transformation by Example.}
Many tools infer migration rules from examples~\cite{DBLP:conf/kbse/AndersenL08, pyevolve, lase, apifix, apimigrator, inferrule, a3, meditor}. Most of these techniques assume access to projects that were manually migrated from one library to another, but such examples are rare~\cite{studyJava}. Finding them typically requires mining large repositories and reconstructing change histories~\cite{islam2024characterizing}. \tool can complement these approaches by automatically distilling code migration knowledge from LLMs into examples that can serve as training data for such tools, reducing the need to rely on manually curated migration histories.

Only recently have these tools begun to leverage production-ready transformation engines like \melt~\cite{melt}, \texttt{PyEvolve}~\cite{pyevolve}, and \texttt{TCInfer}~\cite{tcinfer, inferrule}, which express migration logic in the \texttt{comby} language. Among them, \melt is the most closely related to our work: it targets breaking changes, while \tool focuses on cross-library migrations. We use \melt’s inference algorithm to generate an initial rule set, which we then refine using our agentic approach to produce complete migration scripts in \piranha. Our method could also benefit from recent work on transformation rule generalization, particularly \texttt{Pycraft}~\cite{dilhara2024unprecedented}, which generates variants of individual rules for broader applicability, addressing a known limitation of both \piranha and \texttt{comby} (i.e., how to generalize rules beyond the specific example they were inferred from).

\vspace{1ex}
\noindent\textbf{LLMs for Code Transformation.}
Recent developments show growing interest in using LLMs for automating code migration. Industry tools such as \texttt{Amazon Q Developer}\cite{AmazonQTransform} and reports of LLM-assisted Java upgrades\cite{ziftci2025migrating} reflect this trend. Academic efforts include \texttt{DeepMig}\cite{di2025deepmig} and Islam et al.\cite{islam2025using}. \texttt{DeepMig} relies on supervised training over migration pairs, whereas \tool extracts its own examples without requiring mining corpora.

\texttt{SOAR}~\cite{soar} is one of the earliest approaches to leverage language models for migration. It uses API documentation to infer correspondences between functions in the source and target libraries, then used to migrate client code via program synthesis. Unlike \tool, it does not generate migration examples or transformation scripts. Both approaches use language models to capture migration knowledge: \tool through prompting and pretraining, \texttt{SOAR} through embedding similarity.
\melt~\cite{melt} uses a pull request that introduces a breaking change to prompt an LLM to generate transition examples from the old to the new version, combining model knowledge with example code and tests. \tool uses a more elaborate self-instruction loop to generate \ideas, implementations, and migrations, extending \melt’s approach to a broader domain.

\section{Conclusion}
\label{sec:conclusion}

%Library migration is a common but labor-intensive task in software development. Developers often need to adapt code as dependencies evolve, a process that is tedious, error-prone, and difficult to scale. We present \tool, a system that extracts latent migration knowledge from large language models and converts it into structured, reusable edits expressed in \piranha.
Library migration is a frequent, labor-intensive task for developers. We present \tool, a system that extracts migration knowledge from large language models and converts it into reusable \piranha scripts.
%Instead of relying on manually migrated corpora, \tool prompts LLMs to generate paired implementations and tests for a source and target library. It then distills generalized transformations through a combination of anti-unification and agentic synthesis. The resulting scripts are compatible with modern refactoring workflows and can be inspected, versioned, and applied across large codebases.
Instead of relying on manually migrated corpora, \tool prompts LLMs to generate code and tests for both source and target libraries, then distills generalized transformations using anti-unification and agentic synthesis. The resulting scripts are compatible with modern refactoring workflows and can be inspected, versioned, and applied across large codebases.

%We evaluated \tool on ten Python migration tasks and generated 870 validated migration examples. Using a single synthesis attempt per example, \tool produced correct scripts for 61.6\% of cases and successfully migrated 63.3\% of sibling implementations. It also outperformed MELT, a prior rule inference system, on every task. When applied to 18 real-world open-source repositories, \tool’s scripts performed hundreds of edits and preserved test behavior in the majority of cases. These results show that LLMs encode actionable migration knowledge that can be extracted, validated, and formalized into maintainable transformation logic.
We evaluated \tool on ten Python migration tasks, generating 870 validated migration examples. With one synthesis attempt per example, \tool produced correct scripts for 61.6\% of cases and migrated 63.3\% of sibling implementations, consistently outperforming MELT. Applied to 18 real-world open-source repositories, \tool made hundreds of edits while preserving test behavior in most cases. These results show that LLMs encode actionable migration knowledge that can be extracted, validated, and formalized into maintainable transformation logic.

\bibliographystyle{ACM-Reference-Format}
\bibliography{references}

\end{document}